\documentclass[aps,pre,twocolumn,superscriptaddress,groupedaddress]{revtex4}  

\usepackage{graphicx}
\usepackage{epsfig}
\usepackage{bm}
\usepackage{amsmath}
\usepackage{amssymb}
\usepackage{wasysym}
\usepackage{epstopdf}
\usepackage{color}
\usepackage{xcolor}
\usepackage{epstopdf}
\usepackage{upgreek}
\usepackage{natbib}
\usepackage{arydshln}
\usepackage[T1]{fontenc}

\usepackage[colorlinks=false, bookmarks=true]{hyperref}

\newcommand{\etal}{\emph{et al.}}

\begin{document}


\title{Wake-driven Dynamics of Finite-sized Buoyant Spheres in Turbulence}

\author{Varghese Mathai}
\affiliation{ 
Physics of Fluids Group, Faculty of Science and Technology, J. M. Burgers Centre for Fluid Dynamics, University of Twente, P.O. Box 217, 7500 AE Enschede, The Netherlands.}

\author{Vivek N. Prakash}
\affiliation{ 
Physics of Fluids Group, Faculty of Science and Technology, J. M. Burgers Centre for Fluid Dynamics, University of Twente, P.O. Box 217, 7500 AE Enschede, The Netherlands.}
\affiliation{Department of Bioengineering, Stanford University, Stanford, California 94305, USA.}

\author{Jon Brons\footnote{Present address: Department of Mathematics and Physics, \\ Faculty of Engineering and Computing, Coventry University, United Kingdom}}
\affiliation{ 
Physics of Fluids Group, Faculty of Science and Technology, J. M. Burgers Centre for Fluid Dynamics, University of Twente, P.O. Box 217, 7500 AE Enschede, The Netherlands.}

\author{Chao Sun\footnote{Corresponding author: c.sun@utwente.nl}}
\affiliation{ 
Physics of Fluids Group, Faculty of Science and Technology, J. M. Burgers Centre for Fluid Dynamics, University of Twente, P.O. Box 217, 7500 AE Enschede, The Netherlands.}
\affiliation{Center for Combustion Energy and Department of Thermal Engineering, 
Tsinghua University, 100084 Beijing, China.}

\author{Detlef Lohse}
\affiliation{ 
Physics of Fluids Group, Faculty of Science and Technology, J. M. Burgers Centre for Fluid Dynamics, University of Twente, P.O. Box 217, 7500 AE Enschede, The Netherlands.}

\date{\today}
 
\begin{abstract}

Particles suspended in turbulent flows are affected by the turbulence and at the same time act back on the flow. The resulting coupling can give rise to rich variability in their dynamics. 
Here we report experimental results from an investigation of \textcolor{black}{finite-sized buoyant spheres} in turbulence.
We find that even a marginal reduction in the particle's density from that of the fluid can result in strong modification of its dynamics.
\textcolor{black}{In contrast to classical spatial filtering arguments and predictions of particle models, we find that the particle acceleration variance increases with size. We trace this reversed trend back to the growing contribution from wake-induced forces, unaccounted for in current particle models in turbulence.}
Our findings highlight the need for improved multi-physics based models that account for particle wake effects for a faithful representation of buoyant-sphere dynamics in turbulence.

\end{abstract}
 
\maketitle

%
%
%
%
%
%
%
%

\maketitle

Particulate suspensions in turbulent flows are found in a wide range of natural and industrial settings $-$ typical examples include pollutants dispersed in the atmosphere, droplet suspensions in clouds, air bubbles and plankton distributions in the oceans, and sprays in engine combustion~\cite{brown2009acceleration, calzavarini2009acceleration,toschi2009lagrangian,la2001fluid,bourgoin2014focus}. The behavior of a particle in a flow is intricately linked to several quantities such as the particle's size and shape, its density relative to the carrier fluid, and the flow conditions among others~\cite{elghobashi1994predicting}. For modeling purposes, the equations governing particle motion are often \textcolor{black}{simplified to the case of} a dilute suspension of small rigid spheres in a non-uniform flow~\cite{maxey1983equation,gatignol1983faxen}. In this framework, particle motion in turbulence is described by three fundamental control parameters: the ratio of particle size to dissipative length scale~($\Xi \equiv d_p/\eta$), the particle-fluid density ratio~($\Gamma \equiv \rho_p/\rho_f$), and the Taylor Reynolds number (Re$_{\lambda}$) of the carrier flow.

\textcolor{black}{In many practical situations, particles have a finite-size compared to the dissipative length scales of the flow. Experimental studies have addressed the effects of finite size mainly by using neutrally buoyant finite-size particles in homogeneous and isotropic turbulence~\cite{voth2002measurement, brown2009acceleration,qureshi2007turbulent,homann2010finite, bourgoin2011turbulent}. \textcolor{black}{These studies highlighted certain effects of finite size on the particle's statistical properties, namely a decrease in acceleration variance, an increase in correlation times and a decrease in intermittency in the acceleration PDFs~(on increasing the particle's size). 
All three effects could be interpreted through classical inertial range scaling arguments, which propose spatial filtering due to increasing particle size as the underlying mechanism.}} 
\textcolor{black}{From a modeling perspective, these effects were captured by accounting for flow non-uniformity and spatial filtering at the particle's scale through corrections such as the Fax\'en corrections~(FC) to the force terms~\cite{calzavarini2009acceleration}.}

\begin{figure}[!htbp]
\centering
\vspace{-0.2cm}
\includegraphics[width=0.48\textwidth]{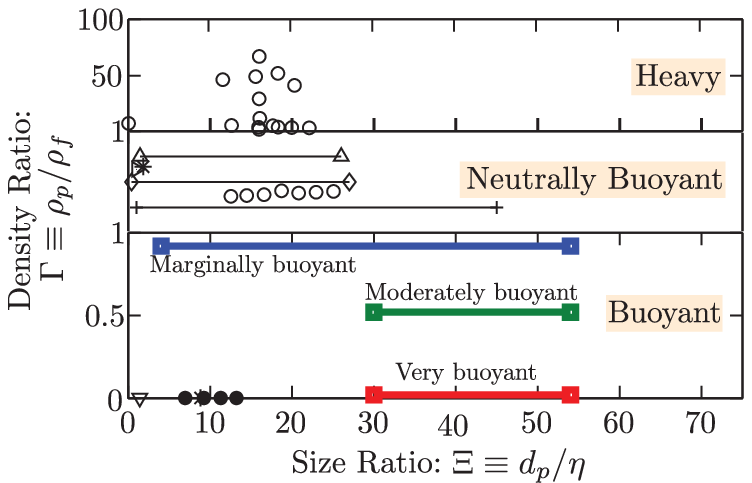}
\vspace{-0.3cm}
\caption{Parameter space of density ratio ($\Gamma$) \textit{vs.} size ratio~($\Xi$) for particles in turbulence. Datapoints from literature:  $\triangle$~-~Voth~\etal~\cite{voth2002measurement}; {$\rhd$}~-~Volk~\etal~\cite{volk2008laser}; {$\ast$}~-~Gibert~\etal~\cite{gibert2010inertial}; {$\diamond$}~-~Brown~\etal~\cite{brown2009acceleration}; {$\circ$}~-~Qureshi~\etal~\cite{qureshi2007turbulent}; +~-~Volk~\etal~\cite{volk2011dynamics}; {$\bigtriangledown$}~-~Mercado~\etal~\cite{mercado2012lagrangian}; {$\CIRCLE$}~-~Prakash~\etal~\cite{prakash2012gravity}; and present experiments: $\textcolor{blue}{\square}$~-~Marginally buoyant, $\textcolor{green}{\square}$~-~Moderately buoyant and $\textcolor{red}{\square}$~-~Very buoyant particles.}
\label{fig:phasediagram}
\end{figure}

\textcolor{black}{The FC model worked reasonably well for finite-sized neutrally buoyant particles~\cite{calzavarini2009acceleration, volk2011dynamics, calzavarini2012impact, homann2010finite, volk2008laser}
 in turbulence. This also led to its extension to predict the behavior of other classes of particles, namely heavy and buoyant particles~\cite{calzavarini2009acceleration, volk2008laser, fiabane2012clustering, prakash2012gravity}. \textcolor{black}{Building on these predictions, some generic models have been proposed to predict the \textit{rms} of the acceleration of arbitrary-density finite-size particles~\cite{zaichik2011model}. These extensions, which practically encompass many of the naturally and industrially relevant particle-laden turbulent flows (where $\Gamma \neq 1$, $\Xi$~>~1), are urgently pending experimental validation.} Since fully resolved numerical simulations~(e.g. Physalis~\cite{naso2010interaction} or Front tracking~\cite{unverdi1992front}) are too expensive for high Re$_{\lambda}$, one is in need of experiments in this regime. However, conducting experiments with non-neutrally buoyant particles has been a challenging task.}
In zero mean-flow turbulence setups~\cite{brown2009acceleration, zimmermann2010lagrangian,chang2012experimental,zocchi1994measurement} these particles would \textcolor{black}{drift} vertically past the small fixed measurement volumes, making it extremely difficult to obtain long particle trajectories in the Lagrangian frame. 

In this \textit{Letter}, we present a novel experimental strategy, wherein a mean-flow may counteract the \textcolor{black}{drift} of the particles. Our investigation covers the regime of finite-sized buoyant spheres (4~$\leq$~$\Xi$~$\leq$~50) in turbulence~(see Fig.~\ref{fig:phasediagram}). The size-ratio ($\Xi$) is defined as the ratio between particle diameter~$d_p$ and the Kolmogorov length scale~$\eta$ in the flow. We study three density ratios~($\Gamma \approx$~0.92,~0.52~\&~0.02), where $\Gamma$ is the ratio of particle density~$\rho_p$ to fluid density~$\rho_f$. According to the $\Gamma$ value, we call the particles either~\textit{marginally buoyant}~($\Gamma \approx$~0.92) or \textit{moderately buoyant}~($\Gamma$~$\approx$~0.52) or \textit{very buoyant}~($\Gamma \approx$~0.02).
The Galileo number, Ga~=~$\sqrt{gd_p^3(1-\Gamma)}/\nu$, provides a good estimate of the \textcolor{black}{buoyancy force in comparison to viscous force}. 
We cover two orders of magnitude variation in Ga~($\approx~30-3000$), and characterize the Lagrangian dynamics of \textcolor{black}{buoyant spheres} in turbulence.



The experiments were performed in the Twente Water Tunnel~(TWT) facility, in which an active grid generated nearly homogeneous and isotropic turbulence in the measurement section~\cite{mercado2012lagrangian,poorte2002experiments}. The water tunnel was configured to have downward flow in the measurement section, and the \textcolor{black}{Taylor Reynolds number of the flow, Re$_{\lambda}$}, was varied from 180 to 300. A small number of rigid buoyant spheres~(0.8~mm~$\leq$~$d_p$~$\leq$~10~mm) were dispersed in each experiment; \textcolor{black}{the volume fraction $\phi$} was kept sufficiently low such that the interaction between the spheres was negligible \textcolor{black}{($\phi \sim O(10^{-5}$))}. The closed circuit enabled the suspended particles to reappear regularly, and hence, sufficient statistics could be obtained~(\textcolor{black}{see Fig.~1 in the supplementary material}~\cite{supplementalmovies}).


\begin{figure}[!htbp]
\centering
\includegraphics[width=0.48\textwidth]{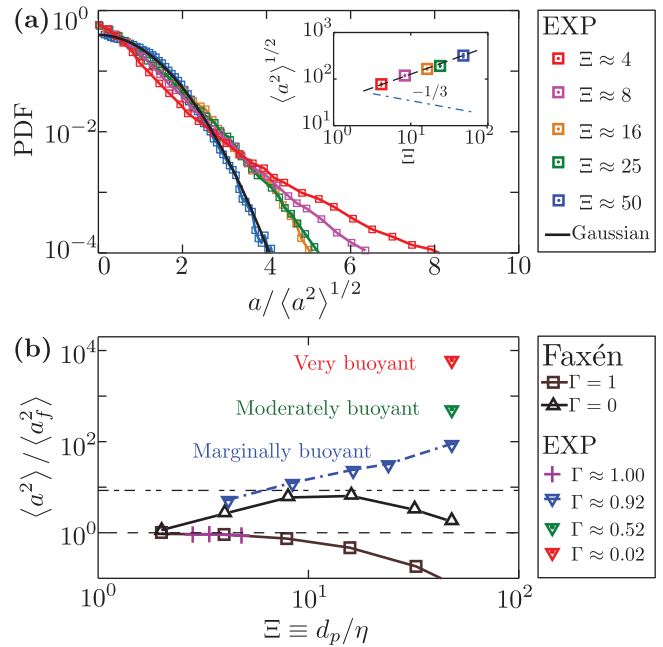}
\caption{Acceleration statistics for buoyant particles in turbulence at Re$_{\lambda} \approx 180$. (a) Acceleration PDF for \textit{marginally buoyant} spheres. Inset shows $\left<a^2\right>^{1/2}$~(mm/$s^{2}$)~\textit{vs.}~$\Xi$~$\equiv$~$d_p/\eta$. \textcolor{black}{The dash-dotted line~(in blue) with slope $\approx$ -1/3 shows the prediction based on classical inertial scaling arguments~\cite{qureshi2007turbulent}.} 
(b)~Normalized acceleration variance from experiments (EXP) compared to the results from Fax\'en-corrected (FC) simulations at Re$_\lambda$~=~180~\cite{calzavarini2009acceleration}. \textcolor{black}{The horizontal dashed and dash-dotted lines~(in black) are lines marking the tracer particle acceleration~($\left<a^2\right>/\left<a_f^2\right> $~=~1) and the upper bound of FC simulations~($\left<a^2\right>/\left<a_f^2\right> $~=~9) respectively. Here, $a_f$ is the fluid tracer acceleration.}}
\label{fig:sizeeffect}
\vspace{-.5cm}
\end{figure}


The particles were imaged using two high speed cameras placed at a 90 degree angle between them. The spheres appear as dark circles in the back-lit images and their diameters corresponded to around 10 pixels. The measurement window size was adjusted to ensure this resolution, and the circle-centers were accurately detected using the Circular~Hough~Transform method. A particle tracking code was used to obtain trajectories of the spheres. These were further subjected to smoothing using a spline based technique~\cite{truscott2012unsteady}, which yielded robust results across the different experimental conditions. Three dimensional trajectories were subsequently obtained by matching  the particle tracks from the individual cameras using a cross-correlation based method.


We first address the question of how a marginal reduction in density ratio ($\Gamma \approx 0.92$) affects the dynamics of finite-sized particles in turbulence. We focus on one of the horizontal components, as in this direction, there is no gravity-induced directional preference. 
In Fig.~\ref{fig:sizeeffect}(a), we show the normalized acceleration probability density function (PDF) for \textit{marginally buoyant}~(MB) spheres at different $\Xi$. The Re$_{\lambda}$ is maintained constant ($\approx 180$), and the particle size ranges from a few Kolmogorov lengths to a fraction of the integral scale~\cite{ishihara2009study}.
At the smallest size ratio ($\Xi \approx 4$) the PDF has wide tails, while it is nearly Gaussian at $\Xi \approx 50$. 
\textcolor{black}{This trend of narrowing PDF tails with increasing $\Xi$ was predicted for buoyant particles using the FC model~\cite{calzavarini2009acceleration, prakash2012gravity}.} Similar behaviour was reported experimentally and also predicted numerically~(FC model) for \textit{neutrally buoyant} spheres in turbulence~\cite{volk2011dynamics,calzavarini2009acceleration}. 

\textcolor{black}{While it seems that accounting for flow non-uniformity and spatial filtering are sufficient to capture the PDF trends, the absolute accelerations reveal something contrary to expectation.}
The inset to Fig.~\ref{fig:sizeeffect}(a) shows that the $rms$ of particle acceleration ($\left<a^2\right>^{1/2}$) increases with $\Xi$, suggesting that larger particles experience greater acceleration fluctuations. \textcolor{black}{This observation is in contrast with both classical inertial range scaling predictions (\textcolor{black}{$\left<a^2\right>^{1/2} \sim \Xi^{-1/3}$}) and the predictions of FC model~(Fig.~\ref{fig:sizeeffect}(b)), according to which larger particles should experience milder acceleration fluctuations due to spatial filtering effects~\cite{voth2002measurement}}. While this $-1/3$ scaling has been experimentally validated for \textit{neutrally buoyant}~(NB) spheres in turbulence~\cite{voth2002measurement, qureshi2007turbulent,brown2009acceleration}, surprisingly, only a marginal reduction in particle density reverses the trend. \textcolor{black}{Furthermore, the accelerations exceed even the upper bound of the Fax\'en model predictions for buoyant particles~i.e.~$\left<a^2\right>/\left<a_f^2\right> $~>~9~(see \textit{marginally buoyant}~($\Gamma \approx 0.92$) case in Fig.~\ref{fig:sizeeffect}(b))}. \textcolor{black}{Here, $a_f$ is the measured fluid tracer acceleration, which was calculated from particle tracking experiments~(conducted separately) using $\approx$ 20~$\mu$m diameter fluorescent tracer particles~\footnote{Fluorescent orange microspheres, Product ID: UVPMS-BO-1.00 from www.cospheric.com
}.} 

\begin{figure}[!htbp]
\centering
\includegraphics[width=0.48\textwidth]{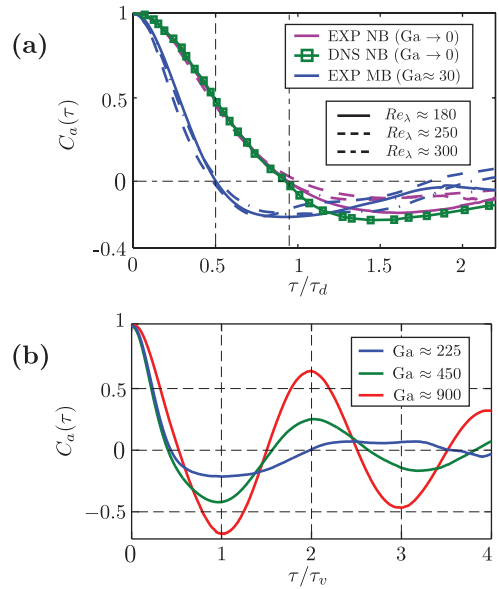}
\caption{(a) Lagrangian acceleration autocorrelation function~$C_a(\tau)$ for marginally buoyant~(MB, $\Gamma$~$\approx$~0.92) and neutrally buoyant~(NB, $\Gamma$~$\approx$~1) spheres with low Ga ($\leq 30$), along with fully resolved DNS predictions from Ref.~\cite{homann2010finite} for NB particles at $Re_{\lambda} = 32$. (b) $C_a(\tau)$ for MB particles~($\Gamma \approx 0.92$) with moderately large Ga ($225 \leq Ga \leq 900$) at Re$_\lambda$~$\approx$~180.
\newline
}
\label{fig:AC}
\vspace{-.5 cm}
\end{figure}

\textcolor{black}{In order to understand the surprisingly large deviations from current particle model predictions, we look into the temporal response of the \textit{marginally buoyant} particles.
The Lagrangian autocorrelation function $C_a(\tau)$ plotted for different particle size-ratios are shown in Fig.~\ref{fig:AC}(a) \& (b). \textcolor{black}{At lower size-ratios (see Fig.~\ref{fig:AC}(a)), the particle accelerations decorrelate according to a turbulent spatial filtering based time scale, $\tau_d~=~(d_p^2/\epsilon)^{1/3}$.} This corresponds to the MB particle with the smallest size-ratio~($\Xi \approx 4$) in the inset to Fig.~\ref{fig:sizeeffect}(a).
\textcolor{black}{At larger size-ratios (see Fig.~\ref{fig:AC}(b), where $\Xi \approx$ 16, 25 and 50), the accelerations decorrelate according to a vortex shedding time-scale, $\tau_v =\frac{d_p}{ St_h \times U_{r}}$, instead of $\tau_d$.} Here, $St_h$ is the Strouhal number and $U_r$ is the measured mean drift velocity of the particle in the turbulent flow.}
This decorrelation behavior is robust across the different Re$_{\lambda}$, $\Xi$, and $\Gamma$ cases (see~table~\ref{table:AC}), with the larger particles additionally displaying strong periodicity. 
\textcolor{black}{We therefore see a gradual transition from the steady drag regime (Ga $\sim$ 30) to a regime with growing vortex shedding-induced effects (Ga from 225 to 900).}

 \textcolor{black}{For a particle in turbulence, there are two effects which can lead to a change in $\left < a^2 \right >^{1/2}$: (i) \textit{the contribution from turbulence} - this decreases with increase in particle size due to spatial filtering effects and (ii) \textit{the contribution from vortex shedding} - this however increases with increasing particle size. Therefore, the total $\left < a^2 \right >^{1/2}$ is the combined effect of turbulence-induced and vortex-shedding-induced forcing. As $\Xi$ increases from 4~to~50~(see inset to Fig.~\ref{fig:sizeeffect}(a)), the unsteady forcing due to vortex shedding starts to outweigh the turbulent forcing at the particle's scale. This results in an overall increase in $\left < a^2 \right >^{1/2}$ with $\Xi$~(for a detailed explanation for the observed trend, see the supplementary material~\cite{supplementalmovies}).} \textcolor{black}{Thus, the agreeing trend of decreasing intermittency in normalised PDFs~(between Fig.~\ref{fig:sizeeffect}(a) and FC model predictions~\cite{calzavarini2009acceleration}) is rather a coincidence, since the absolute particle accelerations are orders of magnitude different.} The present findings reveal the role of wake-driven forces~(vortex~shedding) in buoyant particle dynamics in turbulence.

\begin{table} [!htbp]
\caption{A summary of the explored parameter space. Ga~$-$~Galileo number~(approximate), $\Gamma$~$\equiv$~$\rho_p/\rho_f$~$-$~density ratio, $\Xi$~$\equiv$~$d_p/\eta$~$-$~size ratio, $\tau_D$~$-$~time corresponding to the first minima of $C_a(\tau)$, and $\tau_v$~$-$~vortex shedding timescale based on $St_{h} \approx 0.2$. $Re_{\lambda}$ lies in the range~180$-$300. \textcolor{black}{Supplementary movies A, B, C and D show 3D trajectories of \textit{marginally buoyant}, \textit{moderately buoyant} and \textit{very buoyant} particles~\cite{supplementalmovies}.}} 
\centering
\begin{tabular}{p{0.28\linewidth}p{0.06\linewidth}p{0.1\linewidth}p{0.10\linewidth}p{0.18\linewidth}p{0.16\linewidth}}
\hline
\hline
Particle~type & Ga & $~~~~\Gamma$ & $~~\Xi$ & ${1/\tau_D}$~$(s^{-1})$ & $~~\tau_V/\tau_{D}$\\
\hline
\hline
Very buoyant & 3000 & ~~0.02 & 30-60 & 8.3-8.4 & 1.04$\pm$0.01\\
\hline
Moderately buoyant & 2000 & ~~0.52 & 30-60 & 5.16-5.26 & 0.95$\pm$0.01\\
\hline
~ & 900 & ~~0.92 & 30-60 & 2.45-2.69 & 1.02$\pm$0.05\\
Marginally & 450 & ~~0.92 & 18-36 & 2.5-2.7 &1.04$\pm$0.04 \\
 buoyant & 225 & ~~0.92 & 12-24 & 2.4-2.5 & 0.99$\pm$0.03 \\

~ & 30 & ~~0.92 & ~~4-8 & 4.0-8.0 & 3.00$\pm$1.00\\
\hline
\hline
\end{tabular}
\label{table:AC}
\vspace{.0 cm}

\end{table}

\begin{figure*}[!htbp]
\centering
\includegraphics[width=0.9\textwidth]{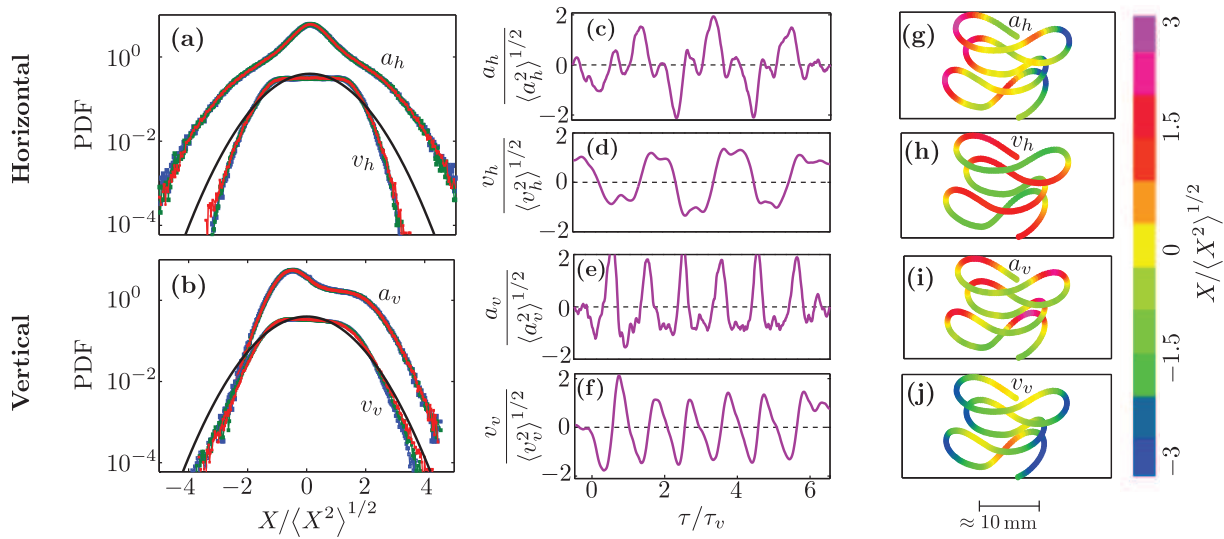}



\caption{PDFs of (a) horizontal and (b) vertical components of velocity and acceleration for a \textit{very~buoyant}~($\Gamma$~$\approx$~0.02) finite-sized sphere at Ga~$\approx$~3000. Blue, green, and red colours correspond to $Re_{\lambda}$ $\approx$ 180, 250, and 300, respectively. \textcolor{black}{The variable $X$ corresponds to $v_h$, $a_h$, $v_y$ and $a_y$ in their respective PDFs. The PDFs have been shifted vertically for clarity in viewing.} A Gaussian profile~(thin~black~line) is overlaid for comparison. (c)-(f) show time traces of velocity and acceleration components of a representative trajectory at Re$_\lambda$~$\approx$~300. (g)-(j)~show the trajectory corresponding to the time traces in (c)-(f), projected on a vertical plane and color-coded with instantaneous quantities.}
\label{fig:Verybuoyant} 
\vspace{-0.5 cm}
\end{figure*}
We now study the effect of the density ratio~$\Gamma$ on buoyant sphere acceleration statistics. In Fig.~\ref{fig:sizeeffect}(b), we show the results of varying particle density at constant~$\Xi$~$\approx$~50. \textcolor{black}{Not surprisingly, the FC model underpredicts the acceleration variance for all three density ratios.} The deviation is greatest for the \textit{`very buoyant'} particle ($\Gamma$~$\approx$~0.02;~Ga~$\approx$~3000), whose experimental result is almost three orders of magnitude higher, suggesting that the particle's accelerations  do not originate from the turbulent forcing. 
In Fig.~\ref{fig:Verybuoyant}(a)~\&~(b),  we present the horizontal and vertical components of the velocity and acceleration PDFs for this \textit{very buoyant} particle at three Re$_{\lambda}$. The horizontal velocity PDF shows a symmetric \textit{flat-head} distribution, and the horizontal acceleration PDF has a \textit{bumped-head} distribution. Both the velocity and acceleration statistics show no observable effect of changes in $Re_{\lambda}$ or $\Xi$. The vertical components have similar characteristics, but the distributions show positive skewness, particularly for acceleration. The Lagrangian autocorrelations of velocity and acceleration are oscillatory (see supplementary material~\cite{supplementalmovies}), with a period comparable to $\tau_v$~(see table \ref{table:AC}). These observations indicate that buoyant particle dynamics is strongly dominated by its wake-induced forces. However, the role of turbulence is still evident in the PDF-tails.

In Fig.~\ref{fig:Verybuoyant}(c)-(f), we plot the time-series of velocity and accelerations of a representative particle trajectory of the \textit{very buoyant} sphere from our experiments. A careful examination reveals that the non-regular PDF shapes have their origin in the time history of the cyclic motions experienced by the particle. In particular, the \textit{flat-headed} velocity, the \textit{bumped-head} acceleration, and the positively skewed vertical PDFs can be traced back to the specific details of the periodic motions.  
In Fig.~\ref{fig:Verybuoyant}(g)-(j), we plot the trajectory coresponding to this time trace projected on a vertical plane.   
The particle traces regular \textit{Lissajous} orbits, only weakly disturbed by the surrounding turbulence~\textcolor{black}{(see supplementary movie~D~\cite{supplementalmovies})}. 
While such periodic motions are classically observed for tethered spheres in laminar flows~\cite{williamson2004vortex,horowitz2010effect}, it is intriguing that these dominate even under highly turbulent conditions, with the turbulence playing only a secondary role.

  
In summary, the dynamics of finite-sized buoyant particles in turbulence can be strongly two-way coupled. 
\textcolor{black}{We show that, despite accounting for finite-size effects, current particle models break down in situations where the particle density is not perfectly matched with the fluid.} 
\textcolor{black}{Wake effects are evident in the experiments, and particle models which ignore these influences are unable to capture the statistics accurately.
Traditional approaches have studied particle dynamics by considering wake-induced~\cite{ern2012wake} and turbulence-induced effects~\cite{toschi2009lagrangian} as separate fields. \textcolor{black}{We emphasize the need to merge these two fields to develop realistic multi-physics based Lagrangian particle models.}} 
\textcolor{black}{Our results provide the first key experimental reference for such emerging models~\cite{verzicco2015multiphysics} in the \textit{particles~in~turbulence} community.}

We gratefully acknowledge A.~Prosperetti and F.~Toschi for inspiring discussions. We thank S.~Huisman for help with the supplementary movies.  We thank X.~Zhu, S.~Wildeman and B.~Gvozdi\'c for comments on the manuscript, and G.-W. Bruggert, M.~Bos and B.~Benschop for technical support. This work was financially supported by the Simon Stevin Prize of the Technology Foundation STW of The Netherlands, and European High-Performance Infrastructures in Turbulence~(EUHIT).


\begin{thebibliography}{33}
\expandafter\ifx\csname natexlab\endcsname\relax\def\natexlab#1{#1}\fi
\expandafter\ifx\csname bibnamefont\endcsname\relax
  \def\bibnamefont#1{#1}\fi
\expandafter\ifx\csname bibfnamefont\endcsname\relax
  \def\bibfnamefont#1{#1}\fi
\expandafter\ifx\csname citenamefont\endcsname\relax
  \def\citenamefont#1{#1}\fi
\expandafter\ifx\csname url\endcsname\relax
  \def\url#1{\texttt{#1}}\fi
\expandafter\ifx\csname urlprefix\endcsname\relax\def\urlprefix{URL }\fi
\providecommand{\bibinfo}[2]{#2}
\providecommand{\eprint}[2][]{\url{#2}}

\bibitem[{\citenamefont{Brown et~al.}(2009)\citenamefont{Brown, Warhaft, and
  Voth}}]{brown2009acceleration}
\bibinfo{author}{\bibfnamefont{R.~D.} \bibnamefont{Brown}},
  \bibinfo{author}{\bibfnamefont{Z.}~\bibnamefont{Warhaft}}, \bibnamefont{and}
  \bibinfo{author}{\bibfnamefont{G.~A.} \bibnamefont{Voth}},
  \bibinfo{journal}{Phys.~Rev.~Lett.} \textbf{\bibinfo{volume}{103}},
  \bibinfo{pages}{194501} (\bibinfo{year}{2009}).

\bibitem[{\citenamefont{Calzavarini et~al.}(2009)\citenamefont{Calzavarini,
  Volk, Bourgoin, Leveque, Pinton, and Toschi}}]{calzavarini2009acceleration}
\bibinfo{author}{\bibfnamefont{E.}~\bibnamefont{Calzavarini}},
  \bibinfo{author}{\bibfnamefont{R.}~\bibnamefont{Volk}},
  \bibinfo{author}{\bibfnamefont{M.}~\bibnamefont{Bourgoin}},
  \bibinfo{author}{\bibfnamefont{E.}~\bibnamefont{Leveque}},
  \bibinfo{author}{\bibfnamefont{J.~F.} \bibnamefont{Pinton}},
  \bibnamefont{and} \bibinfo{author}{\bibfnamefont{F.}~\bibnamefont{Toschi}},
  \bibinfo{journal}{J. Fluid Mech.} \textbf{\bibinfo{volume}{630}},
  \bibinfo{pages}{179} (\bibinfo{year}{2009}).

\bibitem[{\citenamefont{Toschi and Bodenschatz}(2009)}]{toschi2009lagrangian}
\bibinfo{author}{\bibfnamefont{F.}~\bibnamefont{Toschi}} \bibnamefont{and}
  \bibinfo{author}{\bibfnamefont{E.}~\bibnamefont{Bodenschatz}},
  \bibinfo{journal}{Annu. Rev. Fluid Mech.} \textbf{\bibinfo{volume}{41}},
  \bibinfo{pages}{375} (\bibinfo{year}{2009}).

\bibitem[{\citenamefont{La~Porta et~al.}(2001)\citenamefont{La~Porta, Voth,
  Crawford, Alexander, and Bodenschatz}}]{la2001fluid}
\bibinfo{author}{\bibfnamefont{A.}~\bibnamefont{La~Porta}},
  \bibinfo{author}{\bibfnamefont{G.~A.} \bibnamefont{Voth}},
  \bibinfo{author}{\bibfnamefont{A.~M.} \bibnamefont{Crawford}},
  \bibinfo{author}{\bibfnamefont{J.}~\bibnamefont{Alexander}},
  \bibnamefont{and}
  \bibinfo{author}{\bibfnamefont{E.}~\bibnamefont{Bodenschatz}},
  \bibinfo{journal}{Nature} \textbf{\bibinfo{volume}{409}},
  \bibinfo{pages}{1017} (\bibinfo{year}{2001}).

\bibitem[{\citenamefont{Bourgoin and Xu}(2014)}]{bourgoin2014focus}
\bibinfo{author}{\bibfnamefont{M.}~\bibnamefont{Bourgoin}} \bibnamefont{and}
  \bibinfo{author}{\bibfnamefont{H.}~\bibnamefont{Xu}}, \bibinfo{journal}{New
  J. Phys.} \textbf{\bibinfo{volume}{16}}, \bibinfo{pages}{085010}
  (\bibinfo{year}{2014}).

\bibitem[{\citenamefont{Elghobashi}(1994)}]{elghobashi1994predicting}
\bibinfo{author}{\bibfnamefont{S.}~\bibnamefont{Elghobashi}},
  \bibinfo{journal}{Appl. Sci. Res.} \textbf{\bibinfo{volume}{52}},
  \bibinfo{pages}{309} (\bibinfo{year}{1994}).

\bibitem[{\citenamefont{Maxey and Riley}(1983)}]{maxey1983equation}
\bibinfo{author}{\bibfnamefont{M.~R.} \bibnamefont{Maxey}} \bibnamefont{and}
  \bibinfo{author}{\bibfnamefont{J.~J.} \bibnamefont{Riley}},
  \bibinfo{journal}{Phys. Fluids} \textbf{\bibinfo{volume}{26}},
  \bibinfo{pages}{883} (\bibinfo{year}{1983}).

\bibitem[{\citenamefont{Gatignol}(1983)}]{gatignol1983faxen}
\bibinfo{author}{\bibfnamefont{R.}~\bibnamefont{Gatignol}},
  \bibinfo{journal}{J. Mec. Theor. Appl.} \textbf{\bibinfo{volume}{2}},
  \bibinfo{pages}{143} (\bibinfo{year}{1983}).

\bibitem[{\citenamefont{Voth et~al.}(2002)\citenamefont{Voth, la~Porta,
  Crawford, Alexander, and Bodenschatz}}]{voth2002measurement}
\bibinfo{author}{\bibfnamefont{G.~A.} \bibnamefont{Voth}},
  \bibinfo{author}{\bibfnamefont{A.}~\bibnamefont{la~Porta}},
  \bibinfo{author}{\bibfnamefont{A.~M.} \bibnamefont{Crawford}},
  \bibinfo{author}{\bibfnamefont{J.}~\bibnamefont{Alexander}},
  \bibnamefont{and}
  \bibinfo{author}{\bibfnamefont{E.}~\bibnamefont{Bodenschatz}},
  \bibinfo{journal}{J. Fluid Mech.} \textbf{\bibinfo{volume}{469}},
  \bibinfo{pages}{121} (\bibinfo{year}{2002}).

\bibitem[{\citenamefont{Qureshi et~al.}(2007)\citenamefont{Qureshi, Bourgoin,
  Baudet, Cartellier, and Gagne}}]{qureshi2007turbulent}
\bibinfo{author}{\bibfnamefont{N.~M.} \bibnamefont{Qureshi}},
  \bibinfo{author}{\bibfnamefont{M.}~\bibnamefont{Bourgoin}},
  \bibinfo{author}{\bibfnamefont{C.}~\bibnamefont{Baudet}},
  \bibinfo{author}{\bibfnamefont{A.}~\bibnamefont{Cartellier}},
  \bibnamefont{and} \bibinfo{author}{\bibfnamefont{Y.}~\bibnamefont{Gagne}},
  \bibinfo{journal}{Phys. Rev. Lett.} \textbf{\bibinfo{volume}{99}},
  \bibinfo{pages}{184502} (\bibinfo{year}{2007}).

\bibitem[{\citenamefont{Homann and Bec}(2010)}]{homann2010finite}
\bibinfo{author}{\bibfnamefont{H.}~\bibnamefont{Homann}} \bibnamefont{and}
  \bibinfo{author}{\bibfnamefont{J.}~\bibnamefont{Bec}}, \bibinfo{journal}{J.
  Fluid Mech.} \textbf{\bibinfo{volume}{651}}, \bibinfo{pages}{81}
  (\bibinfo{year}{2010}).

\bibitem[{\citenamefont{Bourgoin et~al.}(2011)\citenamefont{Bourgoin, Qureshi,
  Baudet, Cartellier, and Gagne}}]{bourgoin2011turbulent}
\bibinfo{author}{\bibfnamefont{M.}~\bibnamefont{Bourgoin}},
  \bibinfo{author}{\bibfnamefont{N.~M.} \bibnamefont{Qureshi}},
  \bibinfo{author}{\bibfnamefont{C.}~\bibnamefont{Baudet}},
  \bibinfo{author}{\bibfnamefont{A.}~\bibnamefont{Cartellier}},
  \bibnamefont{and} \bibinfo{author}{\bibfnamefont{C.}~\bibnamefont{Gagne}}, in
  \emph{\bibinfo{booktitle}{J. Phys. Conf. Series}} (\bibinfo{organization}{IOP
  Publishing}, \bibinfo{year}{2011}), vol. \bibinfo{volume}{318}, p.
  \bibinfo{pages}{012005}.

\bibitem[{\citenamefont{Volk et~al.}(2008)\citenamefont{Volk, Mordant,
  Verhille, and Pinton}}]{volk2008laser}
\bibinfo{author}{\bibfnamefont{R.}~\bibnamefont{Volk}},
  \bibinfo{author}{\bibfnamefont{N.}~\bibnamefont{Mordant}},
  \bibinfo{author}{\bibfnamefont{G.}~\bibnamefont{Verhille}}, \bibnamefont{and}
  \bibinfo{author}{\bibfnamefont{J.~F.} \bibnamefont{Pinton}},
  \bibinfo{journal}{Europhys.~Lett.} \textbf{\bibinfo{volume}{81}},
  \bibinfo{pages}{34002} (\bibinfo{year}{2008}).

\bibitem[{\citenamefont{Gibert et~al.}(2010)\citenamefont{Gibert, Xu, and
  Bodenschatz}}]{gibert2010inertial}
\bibinfo{author}{\bibfnamefont{M.}~\bibnamefont{Gibert}},
  \bibinfo{author}{\bibfnamefont{H.}~\bibnamefont{Xu}}, \bibnamefont{and}
  \bibinfo{author}{\bibfnamefont{E.}~\bibnamefont{Bodenschatz}},
  \bibinfo{journal}{Europhys. Lett.} \textbf{\bibinfo{volume}{90}},
  \bibinfo{pages}{64005} (\bibinfo{year}{2010}).

\bibitem[{\citenamefont{Volk et~al.}(2011)\citenamefont{Volk, Calzavarini,
  Leveque, and Pinton}}]{volk2011dynamics}
\bibinfo{author}{\bibfnamefont{R.}~\bibnamefont{Volk}},
  \bibinfo{author}{\bibfnamefont{E.}~\bibnamefont{Calzavarini}},
  \bibinfo{author}{\bibfnamefont{E.}~\bibnamefont{Leveque}}, \bibnamefont{and}
  \bibinfo{author}{\bibfnamefont{J.~F.} \bibnamefont{Pinton}},
  \bibinfo{journal}{J.~Fluid Mech.} \textbf{\bibinfo{volume}{668}},
  \bibinfo{pages}{223} (\bibinfo{year}{2011}).

\bibitem[{\citenamefont{Mercado et~al.}(2012)\citenamefont{Mercado, Prakash,
  Tagawa, Sun, and Lohse}}]{mercado2012lagrangian}
\bibinfo{author}{\bibfnamefont{J.~M.} \bibnamefont{Mercado}},
  \bibinfo{author}{\bibfnamefont{V.~N.} \bibnamefont{Prakash}},
  \bibinfo{author}{\bibfnamefont{Y.}~\bibnamefont{Tagawa}},
  \bibinfo{author}{\bibfnamefont{C.}~\bibnamefont{Sun}}, \bibnamefont{and}
  \bibinfo{author}{\bibfnamefont{D.}~\bibnamefont{Lohse}},
  \bibinfo{journal}{Phys. Fluids} \textbf{\bibinfo{volume}{24}},
  \bibinfo{pages}{055106} (\bibinfo{year}{2012}).

\bibitem[{\citenamefont{Prakash et~al.}(2012)\citenamefont{Prakash, Tagawa,
  Calzavarini, Mercado, Toschi, Lohse, and Sun}}]{prakash2012gravity}
\bibinfo{author}{\bibfnamefont{V.~N.} \bibnamefont{Prakash}},
  \bibinfo{author}{\bibfnamefont{Y.}~\bibnamefont{Tagawa}},
  \bibinfo{author}{\bibfnamefont{E.}~\bibnamefont{Calzavarini}},
  \bibinfo{author}{\bibfnamefont{J.~M.} \bibnamefont{Mercado}},
  \bibinfo{author}{\bibfnamefont{F.}~\bibnamefont{Toschi}},
  \bibinfo{author}{\bibfnamefont{D.}~\bibnamefont{Lohse}}, \bibnamefont{and}
  \bibinfo{author}{\bibfnamefont{C.}~\bibnamefont{Sun}}, \bibinfo{journal}{New
  J. Phys.} \textbf{\bibinfo{volume}{14}}, \bibinfo{pages}{105017}
  (\bibinfo{year}{2012}).

\bibitem[{\citenamefont{Calzavarini et~al.}(2012)\citenamefont{Calzavarini,
  Volk, Leveque, Pinton, and Toschi}}]{calzavarini2012impact}
\bibinfo{author}{\bibfnamefont{E.}~\bibnamefont{Calzavarini}},
  \bibinfo{author}{\bibfnamefont{R.}~\bibnamefont{Volk}},
  \bibinfo{author}{\bibfnamefont{E.}~\bibnamefont{Leveque}},
  \bibinfo{author}{\bibfnamefont{J.~F.} \bibnamefont{Pinton}},
  \bibnamefont{and} \bibinfo{author}{\bibfnamefont{F.}~\bibnamefont{Toschi}},
  \bibinfo{journal}{Physica (Amsterdam)} \textbf{\bibinfo{volume}{241D}},
  \bibinfo{pages}{237} (\bibinfo{year}{2012}).

\bibitem[{\citenamefont{Fiabane et~al.}(2012)\citenamefont{Fiabane, Zimmermann,
  Volk, Pinton, and Bourgoin}}]{fiabane2012clustering}
\bibinfo{author}{\bibfnamefont{L.}~\bibnamefont{Fiabane}},
  \bibinfo{author}{\bibfnamefont{R.}~\bibnamefont{Zimmermann}},
  \bibinfo{author}{\bibfnamefont{R.}~\bibnamefont{Volk}},
  \bibinfo{author}{\bibfnamefont{J.-F.} \bibnamefont{Pinton}},
  \bibnamefont{and} \bibinfo{author}{\bibfnamefont{M.}~\bibnamefont{Bourgoin}},
  \bibinfo{journal}{Phys. Rev. E} \textbf{\bibinfo{volume}{86}},
  \bibinfo{pages}{035301} (\bibinfo{year}{2012}).

\bibitem[{\citenamefont{Zaichik and Alipchenkov}(2011)}]{zaichik2011model}
\bibinfo{author}{\bibfnamefont{L.~I.} \bibnamefont{Zaichik}} \bibnamefont{and}
  \bibinfo{author}{\bibfnamefont{V.~M.} \bibnamefont{Alipchenkov}},
  \bibinfo{journal}{Int. J. Mult. Flow} \textbf{\bibinfo{volume}{37}},
  \bibinfo{pages}{236} (\bibinfo{year}{2011}).

\bibitem[{\citenamefont{Naso and Prosperetti}(2010)}]{naso2010interaction}
\bibinfo{author}{\bibfnamefont{A.}~\bibnamefont{Naso}} \bibnamefont{and}
  \bibinfo{author}{\bibfnamefont{A.}~\bibnamefont{Prosperetti}},
  \bibinfo{journal}{New J. Phys.} \textbf{\bibinfo{volume}{12}},
  \bibinfo{pages}{033040} (\bibinfo{year}{2010}).

\bibitem[{\citenamefont{Unverdi and Trygvasson}(1992)}]{unverdi1992front}
\bibinfo{author}{\bibfnamefont{S.}~\bibnamefont{Unverdi}} \bibnamefont{and}
  \bibinfo{author}{\bibfnamefont{G.}~\bibnamefont{Trygvasson}},
  \bibinfo{journal}{J. Comput. Phys.} \textbf{\bibinfo{volume}{100}},
  \bibinfo{pages}{25} (\bibinfo{year}{1992}).

\bibitem[{\citenamefont{Zimmermann et~al.}(2010)\citenamefont{Zimmermann, Xu,
  Gasteuil, Bourgoin, Volk, Pinton, and
  Bodenschatz}}]{zimmermann2010lagrangian}
\bibinfo{author}{\bibfnamefont{R.}~\bibnamefont{Zimmermann}},
  \bibinfo{author}{\bibfnamefont{H.}~\bibnamefont{Xu}},
  \bibinfo{author}{\bibfnamefont{Y.}~\bibnamefont{Gasteuil}},
  \bibinfo{author}{\bibfnamefont{M.}~\bibnamefont{Bourgoin}},
  \bibinfo{author}{\bibfnamefont{R.}~\bibnamefont{Volk}},
  \bibinfo{author}{\bibfnamefont{J.~F.} \bibnamefont{Pinton}},
  \bibnamefont{and}
  \bibinfo{author}{\bibfnamefont{E.}~\bibnamefont{Bodenschatz}},
  \bibinfo{journal}{Rev.~Sci.~Inst.} \textbf{\bibinfo{volume}{81}},
  \bibinfo{pages}{055112} (\bibinfo{year}{2010}).

\bibitem[{\citenamefont{Chang et~al.}(2012)\citenamefont{Chang, Bewley, and
  Bodenschatz}}]{chang2012experimental}
\bibinfo{author}{\bibfnamefont{K.}~\bibnamefont{Chang}},
  \bibinfo{author}{\bibfnamefont{G.~P.} \bibnamefont{Bewley}},
  \bibnamefont{and}
  \bibinfo{author}{\bibfnamefont{E.}~\bibnamefont{Bodenschatz}},
  \bibinfo{journal}{J. Fluid Mech.} \textbf{\bibinfo{volume}{692}},
  \bibinfo{pages}{464} (\bibinfo{year}{2012}).

\bibitem[{\citenamefont{Zocchi et~al.}(1994)\citenamefont{Zocchi, Tabeling,
  Maurer, and Willaime}}]{zocchi1994measurement}
\bibinfo{author}{\bibfnamefont{G.}~\bibnamefont{Zocchi}},
  \bibinfo{author}{\bibfnamefont{P.}~\bibnamefont{Tabeling}},
  \bibinfo{author}{\bibfnamefont{J.}~\bibnamefont{Maurer}}, \bibnamefont{and}
  \bibinfo{author}{\bibfnamefont{H.}~\bibnamefont{Willaime}},
  \bibinfo{journal}{Phys.~Rev.~E} \textbf{\bibinfo{volume}{50}},
  \bibinfo{pages}{3693} (\bibinfo{year}{1994}).

\bibitem[{\citenamefont{Poorte and Biesheuvel}(2002)}]{poorte2002experiments}
\bibinfo{author}{\bibfnamefont{R.}~\bibnamefont{Poorte}} \bibnamefont{and}
  \bibinfo{author}{\bibfnamefont{A.}~\bibnamefont{Biesheuvel}},
  \bibinfo{journal}{J. Fluid Mech.} \textbf{\bibinfo{volume}{461}},
  \bibinfo{pages}{127} (\bibinfo{year}{2002}).

\bibitem[{\citenamefont{See Supplemental Material [url]}(2015)}]{supplementalmovies}
\bibinfo{author}{\bibnamefont{See Supplemental Material at http://link.aps.org/
supplemental/10.1103/PhysRevLett.115.124501, which includes Refs. [28--34], for technical details and movies.}} 


\bibitem[{\citenamefont{See Supplemental Material [url]}(2015)}]{ref1}
\bibinfo{author}{\bibnamefont{[first reference in Supplemental Material not already in Letter].}}

\bibitem[{\citenamefont{See Supplemental Material [url]}(2015)}]{ref5}
\bibinfo{author}{\bibnamefont{[fifth reference in Supplemental Material not already in Letter].}}

\bibitem[{\citenamefont{See Supplemental Material [url]}(2015)}]{ref6}
\bibinfo{author}{\bibnamefont{[sixth reference in Supplemental Material not already in Letter].}}

\bibitem[{\citenamefont{See Supplemental Material [url]}(2015)}]{ref7}
\bibinfo{author}{\bibnamefont{[seventh reference in Supplemental Material not already in Letter].}}

\bibitem[{\citenamefont{See Supplemental Material [url]}(2015)}]{ref8}
\bibinfo{author}{\bibnamefont{[eight reference in Supplemental Material not already in Letter].}}

\bibitem[{\citenamefont{See Supplemental Material [url]}(2015)}]{ref9}
\bibinfo{author}{\bibnamefont{[ninth reference in Supplemental Material not already in Letter].}}

\bibitem[{\citenamefont{See Supplemental Material [url]}(2015)}]{ref11}
\bibinfo{author}{\bibnamefont{[eleventh reference in Supplemental Material not already in Letter].}}


\bibitem[{\citenamefont{Truscott et~al.}(2012)\citenamefont{Truscott, Epps, and
  Techet}}]{truscott2012unsteady}
\bibinfo{author}{\bibfnamefont{T.~T.} \bibnamefont{Truscott}},
  \bibinfo{author}{\bibfnamefont{B.~P.} \bibnamefont{Epps}}, \bibnamefont{and}
  \bibinfo{author}{\bibfnamefont{A.~H.} \bibnamefont{Techet}},
  \bibinfo{journal}{J. Fluid Mech.} \textbf{\bibinfo{volume}{704}},
  \bibinfo{pages}{173} (\bibinfo{year}{2012}).

\bibitem[{\citenamefont{Ishihara et~al.}(2009)\citenamefont{Ishihara, Gotoh,
  and Kaneda}}]{ishihara2009study}
\bibinfo{author}{\bibfnamefont{T.}~\bibnamefont{Ishihara}},
  \bibinfo{author}{\bibfnamefont{T.}~\bibnamefont{Gotoh}}, \bibnamefont{and}
  \bibinfo{author}{\bibfnamefont{Y.}~\bibnamefont{Kaneda}},
  \bibinfo{journal}{Annu. Rev. Fluid Mech.} \textbf{\bibinfo{volume}{41}},
  \bibinfo{pages}{165} (\bibinfo{year}{2009}).

\bibitem[{\citenamefont{Williamson and Govardhan}(2004)}]{williamson2004vortex}
\bibinfo{author}{\bibfnamefont{C.}~\bibnamefont{Williamson}} \bibnamefont{and}
  \bibinfo{author}{\bibfnamefont{R.}~\bibnamefont{Govardhan}},
  \bibinfo{journal}{Annu. Rev. Fluid Mech.} \textbf{\bibinfo{volume}{36}},
  \bibinfo{pages}{413} (\bibinfo{year}{2004}).

\bibitem[{\citenamefont{Horowitz and Williamson}(2010)}]{horowitz2010effect}
\bibinfo{author}{\bibfnamefont{M.}~\bibnamefont{Horowitz}} \bibnamefont{and}
  \bibinfo{author}{\bibfnamefont{C.}~\bibnamefont{Williamson}},
  \bibinfo{journal}{J. Fluid Mech.} \textbf{\bibinfo{volume}{651}},
  \bibinfo{pages}{251} (\bibinfo{year}{2010}).

\bibitem[{\citenamefont{Ern et~al.}(2012)\citenamefont{Ern, Risso, Fabre, and
  Magnaudet}}]{ern2012wake}
\bibinfo{author}{\bibfnamefont{P.}~\bibnamefont{Ern}},
  \bibinfo{author}{\bibfnamefont{F.}~\bibnamefont{Risso}},
  \bibinfo{author}{\bibfnamefont{D.}~\bibnamefont{Fabre}}, \bibnamefont{and}
  \bibinfo{author}{\bibfnamefont{J.}~\bibnamefont{Magnaudet}},
  \bibinfo{journal}{Annu.~Rev.~Fluid~Mech.} \textbf{\bibinfo{volume}{44}},
  \bibinfo{pages}{97} (\bibinfo{year}{2012}).

\bibitem[{\citenamefont{Verzicco}(2015)}]{verzicco2015multiphysics}
\bibinfo{author}{\bibfnamefont{R.}~\bibnamefont{Verzicco}},
  \bibinfo{journal}{IUTAM Symp. Dynamics Bubbly Flows}
  \textbf{\bibinfo{volume}{1}}, \bibinfo{pages}{39} (\bibinfo{year}{2015}).

\end{thebibliography}
\end{document}